\documentclass[11pt]{article}
\usepackage{graphicx}

\usepackage[margin=1in]{geometry}

\usepackage{amssymb,amsfonts,amsmath}
\usepackage{epstopdf}
\usepackage{epsfig}
\usepackage[round,numbers,sort&compress]{natbib} % for PNAS
\usepackage{enumerate}
\usepackage{paralist} 
\usepackage{verbatim}
\usepackage{subfig}
\usepackage{authblk}
\usepackage{setspace}

\newcommand{\td}[1]{\stackrel{\cdot}{#1}}
\newcommand{\logit}{\operatorname{logit}}

\begin{document}

%% For titles, only capitalize the first letter
%% \title{Almost sharp fronts for the surface quasi-geostrophic equation}

\title{Escaping the poverty trap: modeling the interplay between economic growth and the ecology of infectious disease}
\author[1]{Georg M. Goerg}
\author[2]{Oscar Patterson-Lomba}
\author[3]{Laurent H\'ebert-Dufresne}
\author[4,$\dagger$]{Benjamin M. Althouse}
\affil[1]{Carnegie Mellon University, Department of Statistics, Pittsburgh, PA 15213\footnote{Present address: Google, Inc, New York, NY, 10011}}
\affil[2]{Mathematical, Computational, and Modeling Sciences Center, 
School of Human Evolution and Social Change, Arizona State University, Tempe, AZ, 85287}
\affil[3]{D\'epartement de Physique, de G\'enie Physique, et d'Optique, Universit\'e Laval, Qu\'ebec (Qu{\'e}bec), Canada}
\affil[4]{Santa Fe Institute, Santa Fe, NM, 87501}

%% The \maketitle command is necessary to build the title page.
\maketitle

\let\oldthefootnote\thefootnote
\renewcommand{\thefootnote}{\fnsymbol{footnote}}
\footnotetext[2]{To whom correspondence should be addressed. Email: \texttt{althouse@santafe.edu}}
\let\thefootnote\oldthefootnote

%%%%%%%%%%%%%%%%%%%%%%%%%%%%%%%%%%%%%%%%%%%%%%%%%%%%%%%%%%%%%%%%

\begin{abstract} 
The dynamics of economies and infectious disease are inexorably linked: economic well-being influences health (sanitation, nutrition, treatment capacity, etc.) and health influences economic well-being (labor productivity lost to sickness and disease). Often societies are locked into ``poverty traps'' of poor health and poor economy. Here, using a simplified coupled disease-economic model with endogenous capital growth we demonstrate the formation of poverty traps, as well as ways to escape them. We suggest two possible mechanisms of escape both motivated by empirical data: one, through an influx of capital (development aid), and another through changing the percentage of GDP spent on healthcare. We find that a large influx of capital is successful in escaping the poverty trap, but increasing health spending alone is not. Our results demonstrate that escape from a poverty trap may be possible, and carry important policy implications in the world-wide distribution of aid and within-country healthcare spending.
\end{abstract}

\textbf{poverty trap;  economics; infectious disease dynamics; SIRS model; economic growth model;  development aid strategies}

%\doublespacing
\onehalfspacing
%\linenumbers

\section{Introduction}

As economics and health are inexorably linked, implementation of appropriate health-centered public policy decisions is a non-trivial undertaking and carry important consequences for long-term economic growth and the health of societies. On the one hand economic health has a large influence on the physical health of a society (through hospitals, improved infrastructure and sanitation, the ability to grow or import healthier food, etc.), while on the other, economic productivity is dependent upon the overall well-being of workers; e.g. sick people cannot (fully) participate in the labor market~\cite{BerkotwizJohnson74_Health_Labor_Force, Wolff05_Disability_LaborSupply_Bulgaria, Brownetal08_Health_LaborParticipation}. Despite this strong interdependence, few studies have examined directly the interplay between economic growth and the ecology of infectious disease~\cite{Bonds2010, Bonds2010Herd, Plucinski2011, Goenka2007, Goenka2011, Delfino2005}. Many studies have focused on the cost-effectiveness of treatments~\cite{Russell1996, Drummond2005}, but only a handful of studies have focused on other economic aspects such as game-theoretic models of vaccination uptake by individuals~\cite{Bauch2003, Bauch2004, Reluga2006, Galvani2007, Vardavas2007, vanBoven2008, Medlock2009}, and positive and negative externalities imposed by the treatment and evolution of infectious diseases~\cite{Althouse2010, Gersovitz2003, Gersovitz2004, Gersovitz2004Vector, Cook2009, GeoffardPhilipson1997}.

In the present study we present a simple model of economic growth and disease transmission extended from a model developed by Bonds et al. (2010)~\cite{Bonds2010} where they demonstrate how poverty traps can be formed through the interplay of infectious disease and economics. We extend their model to include endogenous economic growth and examine under which scenarios a country could foreseeably escape a poverty trap. Our approach is based on a deterministic Susceptible, Infectious, Recovered, Susceptible (SIRS) model~\cite{AndersonMay92_infectious_diseases_humans} coupled with a Solow economic growth model~\cite{Solow56}, allowing economic growth to be endogenous. We parameterize our model with empirical socio-economic data and test two hypotheses for escaping a poverty trap formed by infectious diseases: one, that a large influx of development aid (capital) will increase the absolute amount of money available for healthcare spending allowing disease to be reduced to a level which will allow a economic growth; and two, that increasing the percentage of GDP spent on healthcare (without the addition of development aid) will reduce disease to levels which allow economic growth.

\section{How Can We Improve the Health and Wealth of a Society?}
\label{sec:motivation}

\begin{figure}[t]
\begin{center}
\includegraphics[]{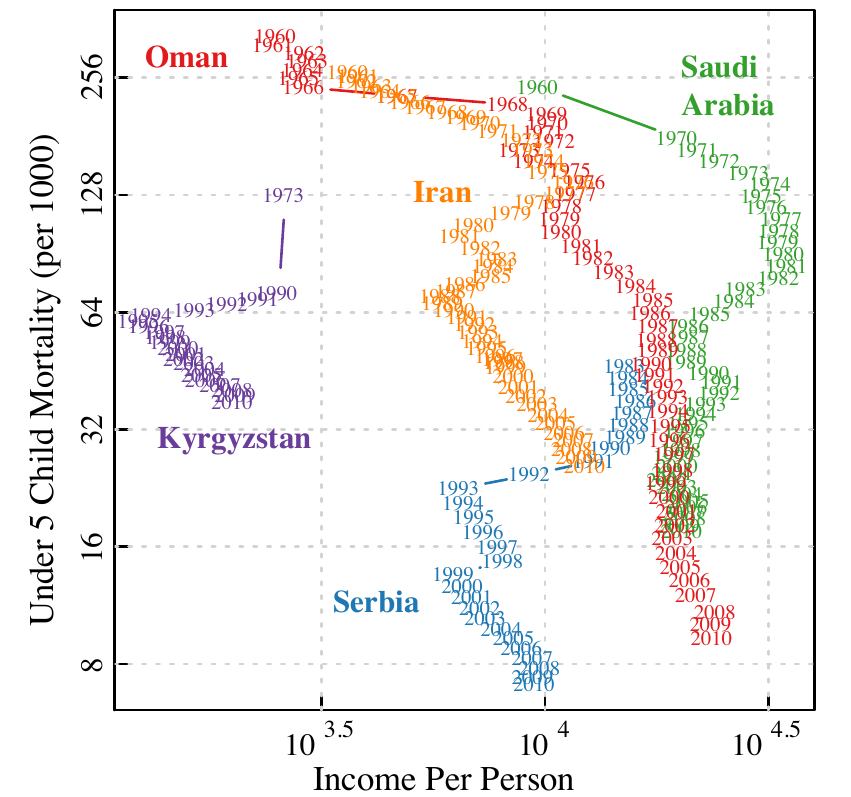} % width = 0.45\textwidth
\caption{\label{fig:World_GDP_child_mortality} Historical evolution of GDP per capita and child mortality for Middle Eastern countries, Kyrgyzstan and Serbia. The exportation of oil in the early sixties caused a large influx of capital to the Middle Eastern countries, after which  child mortality improved substantially. In contrast, Kyrgyzstan and Serbia had a large contraction of available capital but saw improvements in child mortality. Data sources: \texttt{www.gapminder.org} and \texttt{data.worldbank.com}.}
\end{center}
\end{figure}

Figure \ref{fig:World_GDP_child_mortality} illustrates two possible routes out of an unhealthy and poor status and are chosen specifically to motivate our two hypotheses. Child mortality was high in the Middle Eastern countries Oman, Iran and Saudi Arabia in the early nineteen sixties. After the cost of oil rose, their GDP per capita greatly increased, and child mortality continually decreased. Thus, it is possible that a large influx of capital caused an {\em absolute} increase in healthcare spending which thereby improved infrastructure and sanitation (note that Qatar established specialized hospitals in $1979$ with pregnancy care and vaccination programs. Source: \texttt{www.sch.gov.qa/sch/En/scontent.jsp?smenuId=40}), moving these countries from a rich and unhealthy to a rich and healthier society today.

Another route from poor and unhealthy to wealthy and well is through increasing {\em relative} healthcare spending (as a fraction of GDP). This is motivated by Serbia and the former Soviet republic of Kyrgyzstan. After the Soviet Union collapsed, per capita GDP shrank, but child mortality decreased. This may be due to increasing percentage of GDP spent on health. In $1996$ Kyrgyzstan adopted the Manas program, a national primary-care health system and in 1997 implemented a mandatory insurance program~\cite{Kyrgyzstan}. Both examples of national and individual increases in healthcare spending without increases in capital. Similarly for Serbia, after the breakup of Yugoslavia, a shift towards private insurance and increases in healthcare costs (paid both by government and individual) increased healthcare spending, albeit at the cost of efficiency~\cite{Kunitz2004}. The outcome of this being an increase in the proportion of GDP spent on healthcare and a decline in child mortality.

\begin{figure}[t]
\begin{center}
\includegraphics[]{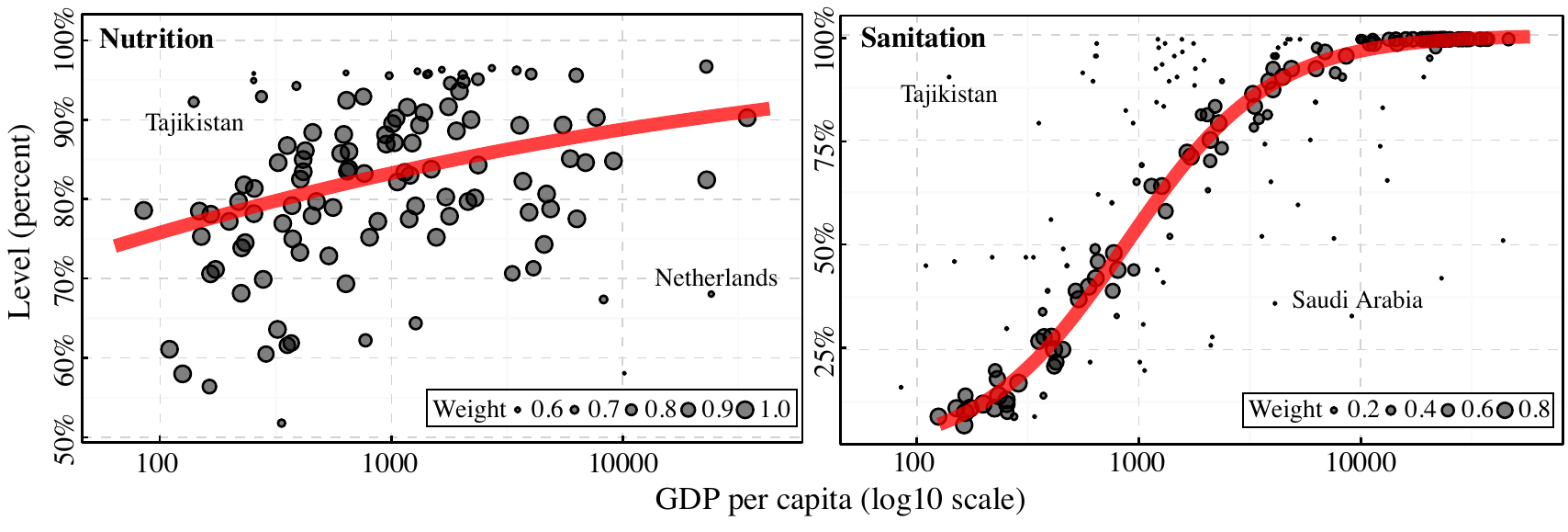} % width = 0.45\textwidth
\caption{\label{fig:logistic_regression_sanitation_nutrition} Weighted logistic regression for sanitation $s(k)$ and nutrition $n(k)$ (note that estimated slope parameters are for logarithm to base $2$ in \eqref{eq:sanitation_log_GDP}). Estimates: Sanitation: $\widehat{\theta}_0^{(s)} = -8.98(0.3)$, $\widehat{\theta}_1^{(s)} = 0.92(0.03)$ for $\log_2(k)$; Nutrition: $\widehat{\theta}_0^{(n)} = 0.21(0.35)$, $\widehat{\theta}_1^{(n)} = 0.14(0.04)$ for $\log_2(k)$.
}
\end{center}
\end{figure}

\section{The Model}
\label{sec:model}

\subsection{Solow's Model of Economic Growth}\label{sec:solow_model}

A classical model of economic growth was proposed by Solow~\cite{Solow56}. Capital, $K$, accumulates by saving a portion, $\sigma \in [0,1]$ of the production, $Y = F(K, L)$, where $F(K,L)$ is the production function for capital $K$ and labor force $L$, and capital decreases due to inflation at rate $\delta$.  Thus, capital changes according to
\begin{equation}
\label{eq:solow_model}
\td{K} = \sigma F(K,L) - \delta K.
\end{equation}

To utilize \eqref{eq:solow_model} we have to specify a functional form for the production function. Here we use the Cobb-Douglas production function~\cite{Houthakker55_CobbDouglas, Goldberger68_EstimationCobbDouglas}:

\begin{equation}
\label{eq:Cobb_Douglas}
F(K,L) = A \left( K^{\alpha} L^{\beta} \right), \quad \alpha, \beta \in [0,1],
\end{equation}

\noindent where $\alpha$ and $\beta$ parametrize capital and labor returns on production, and $A$ is the \emph{multifactor productivity} representing technology and innovation.  An important case of the Cobb-Douglas production function is when $\alpha + \beta = 1$, corresponding to constant returns to scale, i.e.\ doubling $K$ and $L$ also doubles $Y$.

Since our timescale of analysis is relatively short ($<5$ years) we set $\delta = 0$ for simplicity. Furthermore, we restrict our analysis to the basic model \eqref{eq:solow_model} and set $A = 1$, assuming no improvements in innovation and technology.

\subsection{SIRS Model of Disease Transmission}
\label{sec:SIRS_model}

We use a standard SIRS model to describe the dynamics of a generic disease in a population of size $N$ \cite{AndersonMay92_infectious_diseases_humans}.  The population is divided into three non-overlapping compartments: susceptible ($S$), infected ($I$), and recovered ($R$).  By definition, $N = S + I + R$. The dynamics of the system are represented by the following ordinary differential equations (ODEs):

\begin{eqnarray}
\label{eq:S_dot_SIRS}
\dot{S} &=&  (1-\varrho_{p}) \mu N - C \cdot \iota \frac{I}{N} S+\kappa R  -\nu S \\
\label{eq:I_dot_SIRS}
\dot{I} &=& C \cdot \iota  \frac{I}{N} S  	- \gamma ( 1 + \varrho_{t} b ) I - \nu I \\
\label{eq:R_dot_SIRS}
\dot{R} &=&  \varrho_{p} \mu  N +\gamma ( 1 + \varrho_{t} b ) I   -\kappa R  -\nu R,
\end{eqnarray}

\noindent where $\mu$ is the birth rate, $\nu$ is the death rate, $\varrho_{p}$ is the fraction of newborns that receive prophylaxis, $\varrho_{t}$ is the fraction of infected people that get treated, $C$ is the number of contacts per person per unit time,  $\iota$ is the probability of infection given the contact is infected, $\gamma$ is the recovery rate, $b$ is the ``recovery benefit" (percentage) conferred by being treated, and $\kappa$ is the relapse rate at which recovered people lose immunity and thus become again susceptible to the disease.

\begin{figure}[t]
\begin{center}
\includegraphics[]{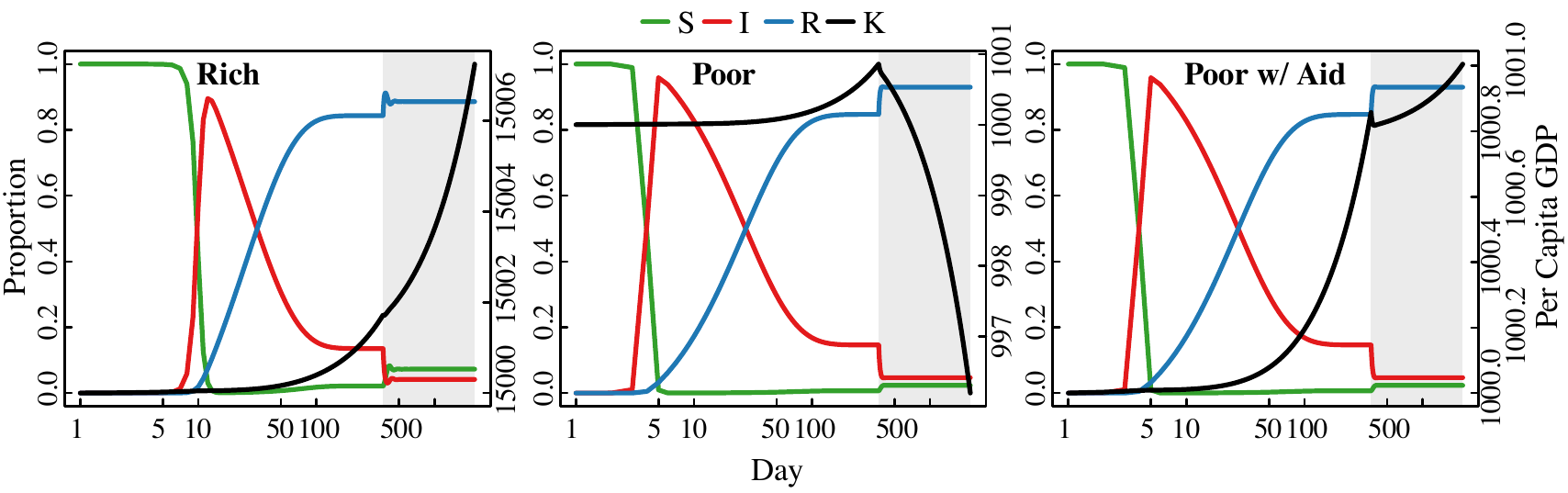}
\caption{\label{fig:SIRTrajectories} Effect of health budget proportion increase from $\phi=0$ to $\phi = 10^{-3}$ at $t = 365$ days for rich (left) and poor (center and right) countries. Development aid changes economic decline (black line, center plot) to economic growth (black line, right plot). Initial conditions: $N = 10^6$, $S = N - 1$, $I = 1$, $R = 0$, and $k = \$ 15 , 000$ (rich) or $k = \$ 1 , 000$ (poor).}
\end{center}
\end{figure}

\subsection{Solow - SIRS Model of Wealth and Health}
\label{sec:solow_SIRS_model}

The impact of health in the Solow model is two-fold: first, infected people cannot participate at their fullest in the labor force (i.e.\ they only have a fraction $\ell \in [0,1]$ of productivity; if infected people cannot work, $\ell = 0$), and second, prophylaxis and treatment incurs costs. Thus, 

\begin{eqnarray}
\label{eq:solow_model_extended_to_SIRS}
\td{K} &=& \underbrace{\sigma \left( K^{\alpha} \left( S + \ell I + R \right)^{1- \alpha} \right)}_{\text{savings per time}} \nonumber \\
 && -  \underbrace{\left( \varrho_{p} c_{p} \mu \left( S + I + R \right) + c_{t} \varrho_{t} I \right)}_{\text {prophylaxis and treatment costs per time}},
\end{eqnarray}

\noindent where $c_{p}$ is the prophylaxis cost per newborn individual ($\$$/person) and $c_{t}$ is the treatment cost per infected individual per unit time ($\$$/person/time).  While birth and death rate depend on wealth, for simplicity we assume constant $\mu$ and $\nu$ over the time-scale of infection.

Let $\phi$ be the proportion of the budget $K$ that is available to spend on health per unit time.  For public policy strategies we make two important assumptions: first, money will only be spent on treatment if all newborns are prophylaxed; and second, if a country has a large enough health budget to cover the cost of prophylaxis for everyone (cost/time = $c_{p} \mu N$) then it will do so, thus prioritizing prophylaxis. The fraction of newborns that receive prophylaxis equals

\begin{eqnarray}
\label{eq:prophylaxis_ratio}
\varrho_{p}(k) &=& \min\left( \frac{\phi K }{c_p \mu N }, \varrho_{p,\max} \right) \\ 
&=& \min\left( \frac{\phi k }{c_p \mu}, \varrho_{p,\max} \right) 
\end{eqnarray}

\noindent where $\varrho_{p, \max}$ is an upper bound on the maximum proportion of people who (choose to) get prophylaxis, and $k = \dfrac{K}{N}$ is GDP  per capita. 

Let $K_p = c_p \varrho_{p}(k) \mu N$ be the total amount of money spent on prophylaxis per unit time ($\$$/time).  Then the remaining health budget can be spent on treatment, and the fraction of infectious individuals being treated equals
\begin{equation}
\label{eq:treatment_ratio}
\varrho_{t} = \min\left( \frac{\phi K - K_p}{c_{t} I}, \varrho_{t,\max} \right),
\end{equation}
where $\varrho_{t, \max}$ is the maximum proportion of infected individuals that are able to be treated. To test the robustness of our findings we also use an entirely opposite strategy where health budget is spent primarily on treatment, and the remaining budget is spent on prophylaxis. Qualitative results remain unchanged (see the Electronic supplementary material [ESM] for details).

The maximum prophylaxis fraction can be inferred from data on immunization indicators, such as vaccination percentage of one-year olds for e.g.\ Hepatitis B, {\em Haemophilus influenzae} B, or Meningococcal vaccine (MCV); all increase with $K$, but plateau at $\varrho_{p, \max} \approx 98\%$. Costs per prophylaxis, $c_p$, can be set to reliable values (ranging from $\approx \$1 - \$5$ USD)~\cite{Wolfson2008}. In the case of treatment, both parameters are highly disease-specific and thus have large variability. Here we assume a cost of $c_t = \$ 50$ USD per treatment per year, with a maximum treatment fraction of $80\%$. To test the impact of changing healthcare spending (as percentage of GDP), we vary $\phi$.

Economic status affects most parameters in the SIRS model. Thus Equations \eqref{eq:S_dot_SIRS} and \eqref{eq:I_dot_SIRS} become

\begin{eqnarray}
\label{eq:SIRS_Sdot_Solow}
\td{S} &=& (1-\varrho_{p}(k)) \mu N - \beta(s(k)) \cdot S \cdot \frac{I}{N} + \kappa(N-S-I) -\nu S \\
\label{eq:SIRS_Idot_Solow}
\td{I} &=& \beta(s(k)) \cdot S \cdot \frac{I}{N} - \gamma(n(k)) (1 + \varrho_{t}(K) b) I - \nu I\\
\label{eq:SIRS_Ndot_Solow}
\td{N} &=& (\mu - \nu)N
\end{eqnarray}

\noindent Here transmissibility, $\beta(s(k)) = C \cdot \iota(s(k))$, and recovery, $\gamma(n(k)) = \tau n(k)$, depend on sanitation $s$ and nutrition $n$, respectively.  The direct dependence of sanitation and nutrition on GDP per capita can be estimated from real-world data. However, due to lack of data on infection probabilities the effect of $s$ on $\iota$ must be chosen on a qualitative basis. We assume that $\iota$ decreases as sanitation increases ($\frac{\partial}{\partial s} \iota(s) < 0$), and that an increase in capital has only small effects on infection for very low and very high sanitation levels, but has large effects on infection at intermediate levels of sanitation (see the ESM).

\section{Data, Modeling Dependence, and Parameter Estimation} 
\label{sec:data_parameter_estimation}

The data shown in Figure \ref{fig:logistic_regression_sanitation_nutrition} were obtained from the World Bank Data Repository (\texttt{data.worldbank.org}) and from \texttt{www.gapminder.org}.  Logistic regression to predict sanitation from GDP per capita not only gives a good fit (Figure \ref{fig:logistic_regression_sanitation_nutrition}), but also passes standard model checks. The functional relationship thus becomes

\begin{equation}
\label{eq:sanitation_log_GDP}
s(k) = 1 - \frac{1}{1 + \exp \left( \theta_0 + \theta_1 \log k \right)} = 1 - \frac{1}{1 + \tilde{\theta}_0 k^{\theta_1}},
\end{equation}

\noindent where $\tilde{\theta}_0 = e^{\theta_0}$. We proceed analogously for nutrition $n(k)$.  Using the $\logit(p) = \dfrac{p}{1-p}$ transform, \eqref{eq:sanitation_log_GDP} becomes linear in $k$
\begin{equation}
\logit s(k) = \theta_0 + \theta_1 \log k,
\end{equation}

\noindent and $(\theta_0, \theta_1)$ can be estimated from the data using ordinary least squares (and variants).  For the SIRS simulations we use robust parameter estimates (\texttt{rlm} in R~\cite{R10}). Details on data sources, parameter estimates, and model fits can be found in the ESM.

\begin{figure}[t]
\begin{center}
\includegraphics[]{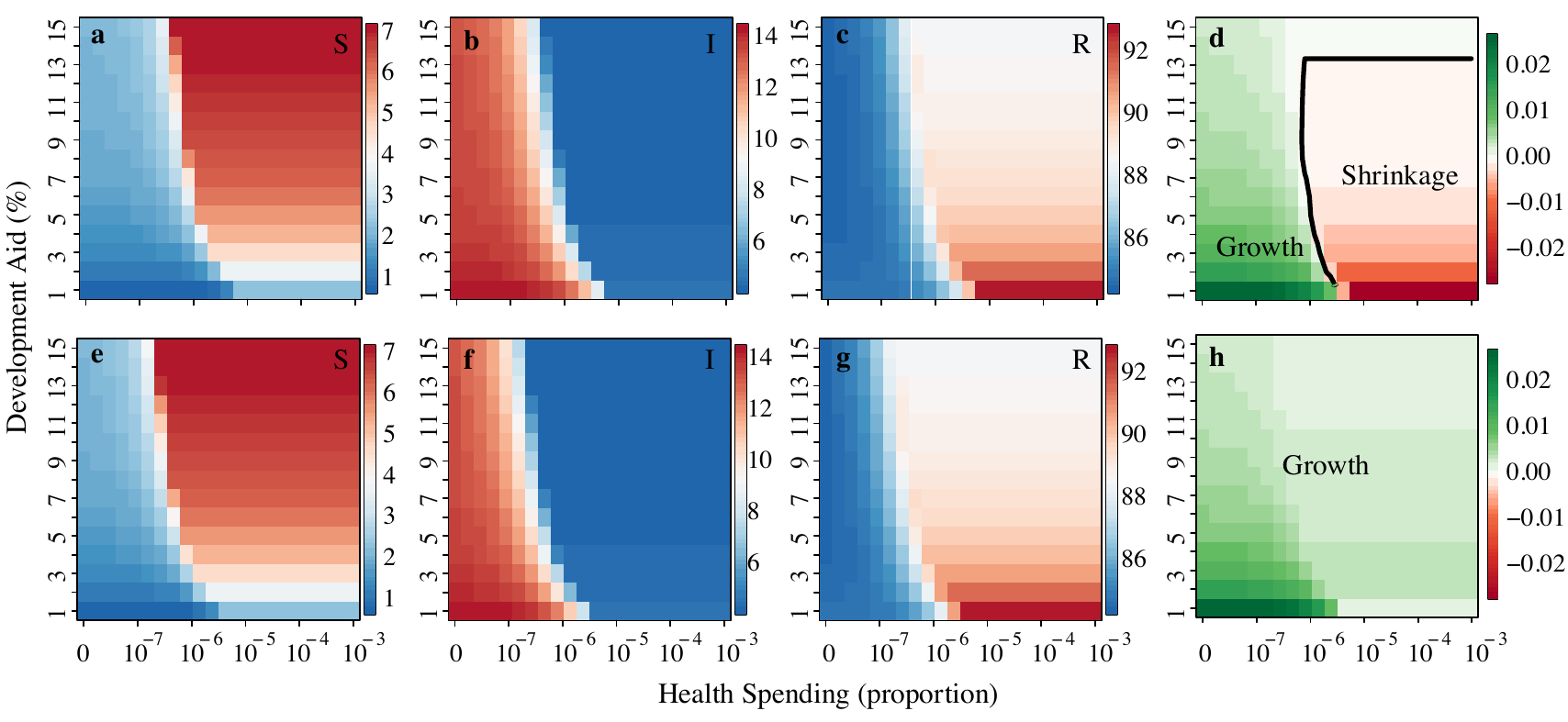}
\caption{\label{fig:sweep_param_SIR_capital_sigma_0_2_alpha_0_2} Phase space diagram of disease transmission and capital growth as a function of $(\phi, $k$)$. SIR endemic proportions (a, b, c, e, f, g) and capital growth (d, h) without (a -- d) and with (e -- h) development aid (reducing costs of prophylaxis and treatment by $50\%$). Larger initial capital $k$ (y-axis) decreases infection, and increasing health share $\phi$ (x-axis) eventually decreases $I$ for all $k$ (a, b, c, e, f, g). Development aid shifts to the left in the value of $\phi$ required to reduce disease prevalence compared to the no aid scenario, i.e. a smaller proportion of per capita healthcare spending is required to achieve the same effects. This is seen in (d) and (h) where without aid, rich economies can increase $\phi$ without drifting into economic decline, while poor countries are caught in the ``poverty trap'' (``Shrinkage'' region), while including aid (cutting costs by $50\%$) helps small $k$ countries to fight infections while still maintaining economic growth. Initial conditions: $N = 10^6$, $S = N - 1$, $I = 1$, $R = 0$, and $c_p= \$ 2$, $c_t = \$ 50$ (no aid).}
\end{center}
\end{figure}

\section{Numerical Results from ODE Simulations}
\label{sec:numerical_results}

For simulations we fix fundamental economic and disease parameters, and consider how different levels of capital ($K$) and health budget policies ($\phi$) affect disease spread as well as economic growth.  Ultimately, policymakers are interested in (a low) infection prevalence and (high) economic growth.

The economy we consider 
\begin{inparaenum}[i)]
\item has a population of one million people ($N = 10^6$),
\item has a savings rate of $\sigma = 0.3$,
\item is labor focused ($\alpha = 0.2$), and
\item has no inflation $\delta = 0$.
\end{inparaenum}
Varying $\sigma$ and $\alpha$ do not make qualitative differences to the main results (see ESM).

As our purpose is to explore two general hypotheses about economic growth and health, we consider a generic disease in a population, {\em i.e.} we wish to keep the analysis from being disease-specific and thus applicable to many pathogens, and thus have chosen parameters to encompass longer-duration infections, such as bacterial pneumonia or parasitic worms~\cite{AndersonMay92_infectious_diseases_humans}. One time step corresponds to 1 day. We assume a
\begin{inparaenum}[i)]
\item maximum recovery rate with a treatment of $\tau = 14.04$ years$^{-1}$,
\item treatment effectiveness of $b = 300\%$ (i.e.\ effective recovery in one week instead of four),
\item birth and death rate of $\mu = \nu = 1/60$ years$^{-1}$,
\item prophylaxis and treatment costs of $c_p = \$2$/person and $c_t = \$50$ per treatment per year,
\item maximum prophylaxis and treatment fraction of $\varrho_{p, \max} = 0.98$ and $\varrho_{t, \max} = 0.8$,
\item $C = 30$ contacts per day,
\item relapse rate of $\kappa = 2$ year$^{-1}$, and 
\item complete disability for work ($\ell = 0$).
\end{inparaenum}
Parameters for sanitation and nutrition as a function of $k$, as estimated, are $\widehat{\theta}_0^{(s)} = -8.98$, $\widehat{\theta}_1^{(s)} = 0.92$, and $\widehat{\theta}_0^{(n)} = 0.21$, $\widehat{\theta}_1^{(n)} = 0.14$, respectively. Calibration parameters for infection probability and recovery rates are given in the ESM.

The numerical solution of the ODEs was obtained for initial conditions: $k(t=0) \in 10^3 \times \lbrace 1, 2, 3, \ldots, 15 \rbrace$ and $\phi = 0$ (no health budget) for $t = 365$ days ($1$ year). At $t=365$ days we raised $\phi$ from zero to $\log_{10} \phi \in \lbrace -8, -7.75, -7.5 \ldots, -3 \rbrace$ and ran the ODEs forward another $365$ time steps.

\subsection{Increasing the Health Budget can Lead into a Poverty Trap}
\label{sec:increasing_health_share}

Figure \ref{fig:SIRTrajectories} shows an example of two extreme economic scenarios. In a ``rich'' country ($k = 15,000$) the proportion of infected people $I$ reaches about $12\%$ after $1$ year and the economy grows exponentially (Figure \ref{fig:SIRTrajectories}, left panel); after the policy change $I$ decreases to about $2\%$ and economic growth slows, but remains positive.  A ``poor'' country ($k = 1,000$) has similar disease prevalence after one year (Figure \ref{fig:SIRTrajectories}, center panel), and increasing the health budget to $\phi = 10^{-3}$ also results in a sharp decline in infections. However these policies lead to negative economic growth, which will consequently lead to more infections in the long run. This is the classic poverty trap as previously described~\cite{Bonds2010}.

Figure \ref{fig:sweep_param_SIR_capital_sigma_0_2_alpha_0_2} shows the phase space plot of $(S,I,R)$ and $\Delta \log k$ (log-returns) at $t=730$ days ($2$ years) as a function of initial capital per capita, $k$, and percentage spent on healthcare, $\phi$. The top-left of Figure \ref{fig:sweep_param_SIR_capital_sigma_0_2_alpha_0_2} show the boundary of effective and economical disease defeat: while both rich and poor economies can lower the proportion of infected people by changing $\phi$, richer countries (higher initial $k$) have to spend proportionally less of their budget on prophylaxis and treatment (smaller $\phi$), than poorer countries do.  Importantly, if a poor country decides to aggressively fight the disease it will temporarily lower prevalence of disease but will end up with negative economic growth (top right, Figure \ref{fig:sweep_param_SIR_capital_sigma_0_2_alpha_0_2}, red region of ``Shrinkage"), ultimately leading to bankruptcy and possibly high prevalence of disease. Rich countries do not fall into this poverty trap due to remaining positive economic growth.

\subsection{Escaping the Poverty Trap}
\label{sec:development_aid}

Countries within poverty traps cannot escape without exogenous input. We aim to test two ways of escaping: one, by increasing capital (development aid) and two, by changing the proportion spent on health care ($\phi$). Figure~\ref{fig:sweep_param_SIR_capital_sigma_0_2_alpha_0_2} (top) shows that once in the poverty trap, increasing $\phi$ will not lead to escape. To explore the other means of escape, we introduce development aid by cutting the cost of prophylaxis and treatment in half ($c_p = \$ 1$ per newborn and $c_t = \$25$ per treatment per year). The right panel of Figure \ref{fig:SIRTrajectories} shows that this cut in costs puts poor countries (initial $k = 1,000$) on a path of economic growth even after increasing $\phi$ (compared to the decline in the center panel).   Similarly, the bottom row of Figure~\ref{fig:sweep_param_SIR_capital_sigma_0_2_alpha_0_2} demonstrates that poor countries can escape the poverty trap and still have a healthy population if they get sufficient development aid. For countries with low savings rates ($\sigma$) or returns on capital ($\alpha$), escape from the poverty trap is possible, albeit at higher levels of development aid (see ESM).

\subsection{How Much Support is Necessary?}

\begin{figure}[t]
\begin{center}
\includegraphics[]{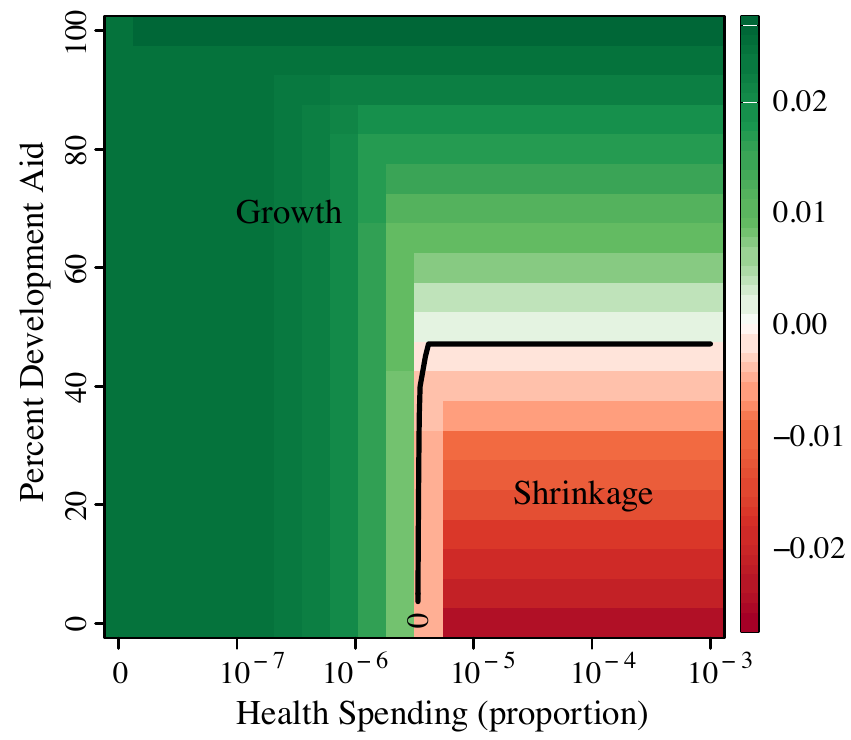} % width = 0.45\textwidth
\caption{\label{fig:sweep_fine_grid_percentage_aid_Capital} How much development aid is necessary? Percentage of development aid (y-axis) versus proportion spent on healthcare, $\phi$, (x-axis) for a country with initial $k = 3,000$. Figure shows capital growth at $t = 730$ days.}
\end{center}
\end{figure}

Development aid in the form of halving the costs of treatment and prophylaxis allow countries to escape the poverty trap, but is this level of support necessary? Figure \ref{fig:sweep_fine_grid_percentage_aid_Capital}  shows the phase space diagram of health spending $\phi$ (x-axis) versus percentage of development aid (y-axis) for a (poor) country with initial $k=3,000$. We see that if development aid exceeds $\approx 50\%$ it can provide long term health and economic benefits for the country. Below this level of aid, even if a country is willing to spend a large share of its budget on health they cannot escape the poverty trap.

\section{Discussion \& Outlook}
\label{sec:discussion}

Economics and the dynamics of infectious disease are deeply linked. Here we extend a model of economics and infectious disease to include endogenous growth of capital. Our aims were twofold: to demonstrate the existence of poverty traps formed by the dynamics of infectious disease (as has been previously demonstrated in~\cite{Bonds2010}) and to test two ways of escaping a poverty trap. Though the framework of our model was relatively simple, we parameterized our model using empirical data in attempts to capture important characteristics of real-world economic and health developments. We found that increasing the percentage of GDP spent on capital alone was not enough to escape the poverty trap, while development aid (through the form of reducing costs of treatment and prophylaxis) allowed countries to reduce disease prevalence and have economic growth, effectively escaping the poverty trap. Our model posits ways to escape poverty traps without having to rely on stochastic effects~\cite{Plucinski2011}.

Although the nature of the current study was qualitative, our parameter estimates were derived from empirical data and our model features endogenous economic growth, to our knowledge, a first in the field. Future work could extend the model in directions such as incorporating innovation and technology growth~\cite{Romer94_EndogenousGrowth}, education as means to increase human capital~\cite{Aghionetal09, Hanushek07_Education_growth, Stoneetal10_EducationEconomicGrowth_RegionPanel}, as well as international trade~\cite{Lucas88}, or migration~\cite{Golgher11_MigrationBrazil}. Future models could include age-structure to capture differential  morbidity and mortality by age. More reliable data to improve parameter estimates, especially nutrition, could increase the realism of the models, as would extending models to incorporate structured populations. 

Determining the optimal level of developmental health aid has been the subject of much debate~\cite{Okamoto2009, Gilson2008}. Our findings underly the importance of sufficient developmental aid in that too little aid (here cutting costs by less than $50\%$) simply fails to pull economies out of spirals of poor health and low wealth. Our model featured a generic disease and could be further extended to focus on a specific malady (e.g.: malaria, diarrheal diseases, HIV, tuberculosis). Many pathogens which affect developing economies are treatable and our results suggest a relative minimum of aid would be necessary to help these countries become healthier and economically productive. Additionally, our model suggests that with aid, countries can lower per capita health care spending and still have economic growth.

While our data-driven model parametrization yields a more realistic interplay between economic and health, our model extension of previous work still paints a largely simplified picture on the decision making process. Our models are not suitable to give quantitative recommendations for policy makers facing specific health hazards. However, we can show that poverty traps can exist if decision makers do not choose their policies wisely. Designing such optimal policies (perhaps through a dynamic optimization framework) remains a task for future research.

A recent report~\cite{LancetCommission13} put forward by a commission of $25$ leading global health experts and economists contends that ``health disparities between nations could be eliminated within a generation'' if funds are invested in healthcare systems of low- and middle-income countries, resulting in enormous social and economic gains. The present work, demonstrating that external aid is key in decreasing the burden of infectious diseases, while also promoting economic growth, offers validating evidence to this report.

\section*{Acknowledgments}
GMG was supported by an INET grant (no.\ IN01100005). Our research team is grateful to NSERC (LHD), WAESOBD  LSAMPBD NSF Cooperative Agreement HRD-1025879 (OPL),  and to Calcul Qu\'ebec for computing facilities. BMA acknowledges the Omidyar Foundation and an NSF Graduate Research Fellowship (grant no.\ DGE-0707427). The authors also wish to thank the Santa Fe Institute and their Complex Systems Summer School at which this work was performed.

\bibliographystyle{unsrtnat}
\bibliography{EconSIRS}	

\begin{thebibliography}{41}
\providecommand{\natexlab}[1]{#1}
\providecommand{\url}[1]{\texttt{#1}}
\expandafter\ifx\csname urlstyle\endcsname\relax
  \providecommand{\doi}[1]{doi: #1}\else
  \providecommand{\doi}{doi: \begingroup \urlstyle{rm}\Url}\fi

\bibitem[Berkowitz and Johnson(1974)]{BerkotwizJohnson74_Health_Labor_Force}
Monroe Berkowitz and William~G. Johnson.
\newblock Health and labor force participation.
\newblock \emph{The Journal of Human Resources}, 9\penalty0 (1):\penalty0 pp.
  117--128, 1974.
\newblock ISSN 0022166X.
\newblock URL \url{http://www.jstor.org/stable/145048}.

\bibitem[Wolff(2005)]{Wolff05_Disability_LaborSupply_Bulgaria}
Fran{\c c}ois-Charles Wolff.
\newblock {Disability and Labour Supply during Economic Transition: Evidence
  from Bulgaria}.
\newblock \emph{LABOUR}, 19\penalty0 (2):\penalty0 303--341, 2005.
\newblock ISSN 1467-9914.
\newblock \doi{10.1111/j.1467-9914.2005.00304.x}.
\newblock URL \url{http://dx.doi.org/10.1111/j.1467-9914.2005.00304.x}.

\bibitem[Brown et~al.(2008)Brown, Roberts, and
  Taylor]{Brownetal08_Health_LaborParticipation}
Sarah Brown, Jenny Roberts, and Karl Taylor.
\newblock Reservation wages, labour market participation and health.
\newblock Working Papers 2008002, The University of Sheffield, Department of
  Economics, 2008.
\newblock URL \url{http://ideas.repec.org/p/shf/wpaper/2008002.html}.

\bibitem[Bonds et~al.(2010)Bonds, Keenan, Rohani, and Sachs]{Bonds2010}
M.H. Bonds, D.C. Keenan, P.~Rohani, and J.D. Sachs.
\newblock Poverty trap formed by the ecology of infectious diseases.
\newblock \emph{Proceedings of the Royal Society B: Biological Sciences},
  277\penalty0 (1685):\penalty0 1185--1192, 2010.

\bibitem[Bonds and Rohani(2010)]{Bonds2010Herd}
Matthew~H Bonds and Pejman Rohani.
\newblock Herd immunity acquired indirectly from interactions between the
  ecology of infectious diseases, demography and economics.
\newblock \emph{J R Soc Interface}, 7\penalty0 (44):\penalty0 541--7, 2010.
\newblock \doi{10.1098/rsif.2009.0281}.

\bibitem[Plucinski et~al.(2011)Plucinski, Ngonghala, and Bonds]{Plucinski2011}
Mateusz~M Plucinski, Calistus~N Ngonghala, and Matthew~H Bonds.
\newblock Health safety nets can break cycles of poverty and disease: a
  stochastic ecological model.
\newblock \emph{J R Soc Interface}, 8\penalty0 (65):\penalty0 1796--803, Dec
  2011.
\newblock \doi{10.1098/rsif.2011.0153}.

\bibitem[Goenka and Liu(2012)]{Goenka2007}
A.~Goenka and L.~Liu.
\newblock Infectious diseases and endogenous fluctuations.
\newblock \emph{Economic Theory}, 50:\penalty0 125--149, 2012.

\bibitem[Goenka et~al.(2011)Goenka, Liu, and Nguyen]{Goenka2011}
Aditya Goenka, Lin Liu, and Manh-Hung Nguyen.
\newblock Infectious diseases and economic growth.
\newblock Working papers, LERNA, University of Toulouse, February 2011.
\newblock URL \url{http://ideas.repec.org/p/ler/wpaper/11.04.338.html}.

\bibitem[Delfino and Simmons(2005)]{Delfino2005}
D.~Delfino and P.J. Simmons.
\newblock Dynamics of tuberculosis and economic growth.
\newblock \emph{Environment and Development Economics}, 10\penalty0
  (6):\penalty0 719, 2005.

\bibitem[Russell et~al.(1996)Russell, Gold, Siegel, Daniels, and
  Weinstein]{Russell1996}
L.B. Russell, M.R. Gold, J.E. Siegel, N.~Daniels, and M.C. Weinstein.
\newblock The role of cost-effectiveness analysis in health and medicine.
\newblock \emph{JAMA: the journal of the American Medical Association},
  276\penalty0 (14):\penalty0 1172--1177, 1996.

\bibitem[Drummond et~al.(2005)Drummond, Sculpher, and Torrance]{Drummond2005}
M.F. Drummond, M.J. Sculpher, and G.W. Torrance.
\newblock \emph{Methods for the economic evaluation of health care programmes}.
\newblock Oxford University Press, USA, 2005.

\bibitem[Bauch et~al.(2003)Bauch, Galvani, and Earn]{Bauch2003}
Chris~T. Bauch, Alison~P. Galvani, and David J.~D. Earn.
\newblock Group interest versus self-interest in smallpox vaccination policy.
\newblock \emph{Proc. Nat. Acad. Sci. USA}, 100\penalty0 (18):\penalty0
  10564--10567, 2003.

\bibitem[Bauch and Earn(2004)]{Bauch2004}
Chris~T. Bauch and David J.~D. Earn.
\newblock Vaccination and the theory of games.
\newblock \emph{Proc. Nat. Acad. Sci. USA}, 101\penalty0 (36):\penalty0
  13391--13394, 2004.

\bibitem[Reluga et~al.(2006)Reluga, Bauch, and Galvani]{Reluga2006}
Timothy~C. Reluga, Chris~T. Bauch, and Alison~P. Galvani.
\newblock Evolving public perceptions and stability in vaccine uptake.
\newblock \emph{Mathematical Biosciences}, 204\penalty0 (2):\penalty0 185--198,
  2006.
\newblock \doi{10.1016/j.mbs.2006.08.015}.
\newblock URL
  \url{http://www.sciencedirect.com/science/article/B6VHX-4KRNR4D-1/2/d114876bb968689217689337b7fb268f}.

\bibitem[Galvani et~al.(2007)Galvani, Reluga, and Chapman]{Galvani2007}
Alison~P. Galvani, Timothy~C. Reluga, and Gretchen~B. Chapman.
\newblock Long-standing influenza vaccination policy is in accord with
  individual self-interest but not with the utilitarian optimum.
\newblock \emph{Proc. Nat. Acad. Sci. USA}, 104\penalty0 (13):\penalty0
  5692--5697, 2007.

\bibitem[Vardavas et~al.(2007)Vardavas, Breban, and Blower]{Vardavas2007}
Raffaele Vardavas, Romulus Breban, and Sally Blower.
\newblock {Can Influenza Epidemics Be Prevented by Voluntary Vaccination?}
\newblock \emph{PLoS Computational Biology}, 3\penalty0 (5):\penalty0 796--802,
  2007.

\bibitem[van Boven et~al.(2008)van Boven, Klinkenberg, Pen, Weissing, and
  Heesterbeek]{vanBoven2008}
Michiel van Boven, Don Klinkenberg, Ido Pen, Franz~J. Weissing, and Hans
  Heesterbeek.
\newblock Self-interest versus group-interest in antiviral control.
\newblock \emph{PLoS ONE}, 3\penalty0 (2):\penalty0 e1558, 2008.
\newblock \doi{10.1371\%2Fjournal.pone.0001558}.
\newblock URL \url{http://dx.doi.org/10.1371\%2Fjournal.pone.0001558}.

\bibitem[Medlock and Galvani(2009)]{Medlock2009}
Jan Medlock and Alison~P Galvani.
\newblock Optimizing influenza vaccine distribution.
\newblock \emph{Science}, 325\penalty0 (5948):\penalty0 1705--8, 2009.
\newblock \doi{10.1126/science.1175570}.

\bibitem[Althouse et~al.(2010)Althouse, Bergstrom, and Bergstrom]{Althouse2010}
Benjamin~M Althouse, Theodore~C Bergstrom, and Carl~T Bergstrom.
\newblock A public choice framework for controlling transmissible and evolving
  diseases.
\newblock \emph{Proc Natl Acad Sci U S A}, 107 Suppl 1:\penalty0 1696--701,
  2010.
\newblock \doi{10.1073/pnas.0906078107}.

\bibitem[Gersovitz and Hammer(2003)]{Gersovitz2003}
Mark Gersovitz and Jeffrey~S. Hammer.
\newblock Infectious diseases, public policy, and the marriage of economics and
  epidemiology.
\newblock \emph{The World Bank Research Observer}, 18\penalty0 (2):\penalty0
  129--157, 2003.

\bibitem[Gersovitz and Hammer(2004{\natexlab{a}})]{Gersovitz2004}
Mark Gersovitz and Jeffrey~S. Hammer.
\newblock The economic control of infectious diseases.
\newblock \emph{The Economic Journal}, 114:\penalty0 1--27, 2004{\natexlab{a}}.

\bibitem[Gersovitz and Hammer(2004{\natexlab{b}})]{Gersovitz2004Vector}
Mark Gersovitz and Jeffrey~S. Hammer.
\newblock Tax/subsidy policies toward vector-borne infectious diseases.
\newblock \emph{Journal of Public Economics}, 89:\penalty0 647--674,
  2004{\natexlab{b}}.

\bibitem[Cook et~al.(2009)Cook, Jeuland, Maskery, Lauria, Sur, Clemens, and
  Whittington]{Cook2009}
Joseph Cook, Marc Jeuland, Brian Maskery, Donald Lauria, Dipika Sur, John
  Clemens, and Dale Whittington.
\newblock Using private demand studies to calculate socially optimal vaccine
  subsidies in developing countries.
\newblock \emph{Journal of Policy Analysis and Management}, 28\penalty0
  (1):\penalty0 6--28, 2009.
\newblock ISSN 0276-8739.
\newblock URL \url{http://www.ncbi.nlm.nih.gov/pubmed/19090047}.
\newblock {PMID:} 19090047.

\bibitem[Geoffard and Philipson(1997)]{GeoffardPhilipson1997}
Pierre-Yves Geoffard and Tomas Philipson.
\newblock Disease eradication: Private versus public vaccination.
\newblock \emph{The American Economic Review}, 87\penalty0 (1):\penalty0
  222--230, 1997.

\bibitem[Anderson and May(1992)]{AndersonMay92_infectious_diseases_humans}
R.M. Anderson and R.M. May.
\newblock \emph{Infectious Diseases of Humans: Dynamics and Control}.
\newblock Oxford Science Publications. Oxford University Press, 1992.
\newblock ISBN 9780198540403.
\newblock URL \url{http://books.google.com/books?id=HT0--xXBguQC}.

\bibitem[Solow(1956)]{Solow56}
Robert~M. Solow.
\newblock A contribution to the theory of economic growth.
\newblock \emph{The Quarterly Journal of Economics}, 70\penalty0 (1):\penalty0
  65--94, 1956.
\newblock ISSN 1531-4650.
\newblock \doi{10.2307/1884513}.
\newblock URL \url{http://dx.doi.org/10.2307/1884513}.

\bibitem[{US Library of Congress}(2007)]{Kyrgyzstan}
{US Library of Congress}.
\newblock Country profile: Kyrgyzstan.
\newblock Technical report, United States Library of Congress, 2007.
\newblock URL \url{http://www.unhcr.org/refworld/docid/46f9134b0.html}.

\bibitem[Kunitz(2004)]{Kunitz2004}
Stephen~J Kunitz.
\newblock The making and breaking of yugoslavia and its impact on health.
\newblock \emph{Am J Public Health}, 94\penalty0 (11):\penalty0 1894--904, Nov
  2004.

\bibitem[Houthakker(1955)]{Houthakker55_CobbDouglas}
H.~S. Houthakker.
\newblock {The Pareto Distribution and the Cobb-Douglas Production Function in
  Activity Analysis}.
\newblock \emph{The Review of Economic Studies}, 23\penalty0 (1):\penalty0 pp.
  27--31, 1955.
\newblock ISSN 00346527.
\newblock URL \url{http://www.jstor.org/stable/2296148}.

\bibitem[Goldberger(1968)]{Goldberger68_EstimationCobbDouglas}
Arthur~S. Goldberger.
\newblock {The Interpretation and Estimation of Cobb-Douglas Functions}.
\newblock \emph{Econometrica}, 36\penalty0 (3/4):\penalty0 464--472, 1968.
\newblock ISSN 00129682.
\newblock URL \url{http://www.jstor.org/stable/1909517}.

\bibitem[Wolfson et~al.(2008)Wolfson, Gasse, Lee-Martin, Lydon, Magan, Tibouti,
  Johns, Hutubessy, Salama, and Okwo-Bele]{Wolfson2008}
Lara~J Wolfson, Fran\c{c}ois Gasse, Shook-Pui Lee-Martin, Patrick Lydon, Ahmed
  Magan, Abdelmajid Tibouti, Benjamin Johns, Raymond Hutubessy, Peter Salama,
  and Jean-Marie Okwo-Bele.
\newblock {Estimating the costs of achieving the WHO-UNICEF Global Immunization
  Vision and Strategy, 2006-2015}.
\newblock \emph{{Bulletin of the World Health Organization}}, 86:\penalty0 27
  -- 39, 01 2008.
\newblock ISSN 0042-9686.

\bibitem[{R Development Core Team}(2010)]{R10}
{R Development Core Team}.
\newblock \emph{R: A Language and Environment for Statistical Computing}.
\newblock R Foundation for Statistical Computing, Vienna, Austria, 2010.
\newblock URL \url{http://www.R-project.org/}.
\newblock {ISBN} 3-900051-07-0.

\bibitem[Romer(1994)]{Romer94_EndogenousGrowth}
Paul~M. Romer.
\newblock The origins of endogenous growth.
\newblock \emph{The Journal of Economic Perspectives}, 8\penalty0 (1):\penalty0
  pp. 3--22, 1994.
\newblock ISSN 08953309.
\newblock URL \url{http://www.jstor.org/stable/2138148}.

\bibitem[Aghion et~al.(2009)Aghion, Boustan, and Hoxby]{Aghionetal09}
P.~Aghion, L.~Boustan, and C.~Hoxby.
\newblock {The Causal Impact of Education on Economic Growth: Evidence from
  U.S.}, 2009.

\bibitem[Hanushek and Woessmann(2007)]{Hanushek07_Education_growth}
Eric~A. Hanushek and Ludger Woessmann.
\newblock The role of education quality for economic growth.
\newblock Policy Research Working Paper Series 4122, The World Bank, 2007.
\newblock URL \url{http://ideas.repec.org/p/wbk/wbrwps/4122.html}.

\bibitem[Stone et~al.(2010)Stone, Bania, and
  Gray]{Stoneetal10_EducationEconomicGrowth_RegionPanel}
Joe Stone, Neil Bania, and Jo~Anna Gray.
\newblock Public infrastructure, education, and economic growth:
  Region-specific complementarity in a half-century panel of states.
\newblock MPRA Paper 21697, University Library of Munich, Germany, 2010.
\newblock URL \url{http://ideas.repec.org/p/pra/mprapa/21697.html}.

\bibitem[Lucas(1988)]{Lucas88}
Robert~E. Lucas.
\newblock On the mechanics of economic development.
\newblock \emph{Journal of Monetary Economics}, 22:\penalty0 3--42, 1988.
\newblock URL
  \url{http://linkinghub.elsevier.com/retrieve/pii/0304393288901687}.

\bibitem[Golgher et~al.(2011)Golgher, De~Figueiredo, and
  Santolin]{Golgher11_MigrationBrazil}
Andr\'e~Braz Golgher, L\'izia De~Figueiredo, and Roberto Santolin.
\newblock {Migration and economic growth in Brazil: empirical applications
  based on the Solow-Swan model}.
\newblock \emph{The Developing Economies}, 49\penalty0 (2):\penalty0 148--170,
  2011.
\newblock ISSN 1746-1049.
\newblock \doi{10.1111/j.1746-1049.2011.00127.x}.
\newblock URL \url{http://dx.doi.org/10.1111/j.1746-1049.2011.00127.x}.

\bibitem[Okamoto et~al.(2009)Okamoto, Nhea, Akashi, Kawaguchi, Ui, Kinoshita,
  and Aoyama]{Okamoto2009}
Miyoko Okamoto, Sithan Nhea, Hidechika Akashi, Leo Kawaguchi, Shiori Ui, Mari
  Kinoshita, and Atsuko Aoyama.
\newblock Developing institutional capacity of health service system management
  at the district level in rural cambodia.
\newblock \emph{Biosci Trends}, 3\penalty0 (6):\penalty0 239--46, Dec 2009.

\bibitem[Gilson and Raphaely(2008)]{Gilson2008}
Lucy Gilson and Nika Raphaely.
\newblock The terrain of health policy analysis in low and middle income
  countries: a review of published literature 1994-2007.
\newblock \emph{Health Policy Plan}, 23\penalty0 (5):\penalty0 294--307, Sep
  2008.
\newblock \doi{10.1093/heapol/czn019}.

\bibitem[Jamison et~al.(2013)Jamison, Summers, Alleyne, Arrow, Berkley,
  Binagwaho, Bustreo, Evans, Feachem, Frenk, Ghosh, Goldie, Guo, Gupta, Horton,
  Kruk, Mahmoud, Mohohlo, Ncube, Pablos-Mendez, Reddy, Saxenian, Soucat,
  Ulltveit-Moe, and Yamey]{LancetCommission13}
Dean~T. Jamison, Lawrence~H. Summers, George Alleyne, Kenneth~J. Arrow, Seth
  Berkley, Agnes Binagwaho, Flavia Bustreo, David Evans, Richard G.~A. Feachem,
  Julio Frenk, Gargee Ghosh, Sue~J. Goldie, Yan Guo, Sanjeev Gupta, Richard
  Horton, Margaret~E. Kruk, Adel Mahmoud, Linah~K. Mohohlo, Mthuli Ncube, Ariel
  Pablos-Mendez, K.~Srinath Reddy, Helen Saxenian, Agnes Soucat, Karene~H.
  Ulltveit-Moe, and Gavin Yamey.
\newblock Global health 2035: a world converging within a generation.
\newblock \emph{The Lancet}, 382\penalty0 (9908):\penalty0 1898--1955, December
  2013.
\newblock ISSN 01406736.
\newblock \doi{10.1016/s0140-6736(13)62105-4}.
\newblock URL \url{http://dx.doi.org/10.1016/s0140-6736(13)62105-4}.

\end{thebibliography}

\end{document}